\documentstyle[epsf,a4]{article}  
\begin{document}                % INITIALIZE - DONT CHANGE
\newcommand{\manual}{rm}        % Substitute rm (Roman) font.
\newcommand\bs{\char '134 }     % add backslash char to \tt font

\newcommand{\simlt}{\stackrel{<}{{}_\sim}}
\newcommand{\simgt}{\stackrel{>}{{}_\sim}}
\newcommand{\MeV}{\;\mathrm{MeV}}
\newcommand{\TeV}{\;\mathrm{TeV}}
\newcommand{\GeV}{\;\mathrm{GeV}}
\newcommand{\eV}{\;\mathrm{eV}}
\newcommand{\cm}{\;\mathrm{cm}}
\newcommand{\s}{\;\mathrm{s}}
\newcommand{\sr}{\;\mathrm{sr}}
\newcommand{\lab}{\mathrm{lab}}
\newcommand{\ts}{\textstyle}
\newcommand{\ol}{\overline}
\newcommand{\be}{\begin{equation}}
\newcommand{\ee}{\end{equation}}
\newcommand{\ba}{\begin{eqnarray}}
\newcommand{\ea}{\end{eqnarray}}
\newcommand{\nn}{\nonumber}
\newcommand{\nm}{{\nu_\mu}}
\newcommand{\pp}{$\overline{p}(p)-p\;\;$}
\renewcommand{\floatpagefraction}{1.}
\renewcommand{\topfraction}{1.}
\renewcommand{\bottomfraction}{1.}
\renewcommand{\textfraction}{0.}               
\renewcommand{\thefootnote}{F\arabic{footnote}}
\title{A new implication for strong interactions if
large, direct CP violation in $\ol{B}^0(B^0)\to \pi^+\pi^-$ is
confirmed}
\author{Saul Barshay and Georg Kreyerhoff\\III. Physikalisches Institut\\
RWTH Aachen\\D-52056 Aachen\\Germany}
\maketitle
\begin{abstract}                % DON'T CHANGE THIS LINE
We show that the large, direct CP-violation parameter $A_{\pi\pi}=-C_{\pi\pi}$,
reported by the BELLE collaboration in the decays  $\ol{B}^0(B^0)\to \pi^+\pi^-$,
implies an unusual situation in which the presence of a very large difference
between two strong-interaction phases ($\sim -110^\circ$) plays an essential role.
We make the demonstration within a model of strong, two-body quasi-elastic interactions
between physical hadrons. The model can accommodate a large difference between
two strong-interaction phases, for which it provides a natural enhancement.
\end{abstract}
Two experiments have very recently published results concerning CP violation
in the decays $\ol{B}^0(B^0)\to \pi^+\pi^-$ \cite{ref1, ref2,ref3}. The results involve
measurements of the two parameters which determine mixing-induced and direct
CP-violation. Direct CP-violation would occur even if the sizable mixing\cite{ref4}
between $B^0$ and $\ol{B}^0$ caused by the known \cite{ref4} mass difference
$\Delta m_d$ between the states $(pB^0\mp q\ol{B}^0)/\sqrt{2}$,
were to vanish (i.~e.~$\Delta m_d\to 0$). When one recalls that more than
25 years passed after the discovery of CP violation\cite{ref5}, before
direct CP violation in the $K^0-\ol{K}^0$ system was established by different
experiments \cite{ref6,ref7,ref8}, it is clear that the significance of the present single
experimental indication \cite{ref1,ref3} for large, direct CP violation in
the $B^0-\ol{B}^0$ system warrants careful consideration. The two general
empirical parameters mentioned above are, of course, related through the
parameters of the standard model and certain dynamical assumptions\cite{ref1,ref2}.
However in the decay mode in which CP violation in the $B^0-\ol{B}^0$ system
was discovered \cite{ref9,ref10}, that due to mixing is large \cite{ref9,ref10},
whereas direct CP violation is not detected \cite{ref9}. It is a general
requirement \cite{ref11,ref12} that significant elastic and/or inelasticity-induced,
strong-interaction phases (at least two different phases) be present,
in order for direct CP violation to be manifest in the decay amplitudes for particle
and antiparticle.\footnote{In 11 is one of the first
published calculations which explicitly demonstrates the necessary role of different
inelasticity-induced, as well as elastic, strong-interaction phases among physical
hadrons in CP violation (sometimes referred to as ``soft'' final-state interactions,
in the ensuing popular literature).} In this paper
we demonstrate within an explicit and clear dynamical model involving
physical hadrons, that if the large, direct CP violation recently reported \cite{ref1,ref3}
is confirmed, then a most unusual situation must hold for the two necessary
strong-interaction phases. This is our main purpose. The model can
accommodate, in principle, the necessary generation of a large difference
between two strong-interaction phases of about $-110^\circ$. It provides a natural
enhancement for the phase difference.

The time evolution of the decays $\ol{B}^0(B^0)\to \pi^+\pi^-$ is
described \cite{ref1,ref2} by \footnote{We use the notation of \cite{ref2}. One must
note that the parameter for direct CP violation $C_{\pi\pi}$, is the same as (-1)
times the parameter denoted by $A_{\pi\pi}$ in \cite{ref1,ref3}.}
\be
f_\pm(\Delta t) = \frac{e^{-|\Delta t|/\tau_{B^0}}}{4\tau_{B^0}}\left\{
1\pm S_{\pi\pi}\sin(\Delta m_d\Delta t)\mp C_{\pi\pi}\cos(\Delta m_d\Delta t)\right\}
\ee
where the upper (lower) sign refers to $\ol{B^0} (B^0)$ decay (with a $B^0 (\ol{B}^0)$
tagged). The phenomenological parameters\cite{ref13} $S_{\pi\pi}$ and
$C_{\pi\pi}$, for mixing-induced and direct CP violation respectively, are
defined by
\renewcommand{\theequation}{2a}
\be
S_{\pi\pi}=\frac{2{\mathrm{Im}}\lambda}{1+|\lambda|^2},\;\;\; C_{\pi\pi}=
\frac{1-|\lambda|^2}{1+|\lambda|^2}
\ee
\par\noindent
with $\lambda=(q/p)(\ol{B}_{\pi\pi}/B_{\pi\pi})$. The states (heavy (H) and
light (L)) with mass difference $\Delta m_d$ \cite{ref13} are 
\renewcommand{\theequation}{2b}
\be
|B_{H,L}> = (p|B^0> \mp q |\ol{B}^0>)/\sqrt{2}; \;\;\; |q/p|\cong 1
\ee
\renewcommand{\theequation}{\arabic{equation}}
\setcounter{equation}{2}
\par\noindent
Here $\ol{B}_{\pi\pi}$ and $B_{\pi\pi}$ denote the complex decay amplitudes
for $\ol{B}^0 (B^0)\to \pi^+\pi^-$, respectively. In the model developed
below, these amplitudes depend upon two strong-interaction phases, and upon
a CP-violating weak-interaction phase; an explicit dependence is exhibited
in Eqs.~(7,8) below. The BELLE\cite{ref1,ref3} and BABAR\cite{ref2} results
are:
\ba
{\mathrm{BELLE}}: && S_{\pi\pi} = -1.23 \pm 0.41({\mathrm{stat}})\;^{+ 0.08}_{-0.07} ({\mathrm{syst}}) \nn\\
                  && C_{\pi\pi} = -A_{\pi\pi} = -0.77 \mp 0.27 ({\mathrm{stat}})\mp 0.08 ({\mathrm{syst}})\nn\\
{\mathrm{BABAR}}:  && S_{\pi\pi} = +0.02 \pm 0.34({\mathrm{stat}}) \pm 0.05 ({\mathrm{syst}}) \nn\\
                  && C_{\pi\pi} = -A_{\pi\pi} = -0.30 \pm 0.25 ({\mathrm{stat}})\pm 0.04 ({\mathrm{syst}})
\ea
As we shall see below, a remarkable feature of the BELLE data is the large
magnitude of $C_{\pi\pi}(-A_{\pi\pi})$, \footnote{The physical boundary is at $(S_{\pi\pi}^2+C_{\pi\pi}^2)=1$.
Approximate formulae from our model and from other models \cite{ref3}, may allow
larger values near to the boundary, for certain values of the parameters. So does the present BELLE data,
presumably from a statistical fluctuation.\cite{ref1,ref3}} 
the parameter which describes direct CP violation (i.~e.~is present for $\Delta m_d\to 0$). 

We begin the analysis with the determination of two strong-interaction eigenstates \footnote{The
consideration of two eigenstates is a minimal idealization which allows one to see clearly
the essential features of the strong and weak-interaction elements in the physics, when the
$\eta\eta_c (c\ol{c})$ particle (quark) content of the mixed system is considered. All three
quark generations come into play, through charged currents.}
which mix the final $\pi^+\pi^-$ system (the component with isospin zero) at a total (c.~m.~)
energy of $m_{B^0} = 5.28 \GeV$, with another two-particle system of distinctly different
particle content, namely $\eta\eta_c$.\cite{ref12} \footnote{One could equally well consider
the system $\eta'\eta_c$. Both systems were used in \cite{ref12}, where strong-interaction
phases were calculated which are relevant for estimating sizable direct CP-violating asymmetries
in the charged decay modes $B^\pm \to (\pi^\pm \eta, \pi^\pm\eta', \pi^\pm \eta_c)$.
A CP-violating asymmetry in a charged-particle decay mode remains to be discovered
(for $B$ or $K$). BELLE may be on the track (private communication from K.~Abe).
} This particular system
is mixed by \underline{quasi-elastic} strong interaction \cite{ref11,ref12} with $\pi\pi$.$^{F5}$
The reaction is quasi-elastic at a c.~m.~energy of 5.28 GeV, the momentum in the $\eta\eta_c$ system
is 0.7 times that in the $\pi\pi$ system. The $c\ol{c}$ component which is implicit in the
constitution of the system$^{F5}$ $\eta\eta_c$, brings the complex phase 
$\delta$ ( or $\delta_{13}$)\cite{ref14} of the CKM matrix (necessary for CP violation in the
standard-model weak interaction) directly into play in the decay amplitudes $\ol{B}_{\pi\pi}$
and $B_{\pi\pi}$ (Eq.~(6) below). Consider a two-by-two strong-interaction $K$ matrix for 
quasi-elastic scattering of the form
\be
K = \begin{array}{cc}  & \begin{array}{cc} \pi\pi\;\;\; &  \;\;\;\eta\eta_c \end{array} \\
                           \begin{array}{c} \pi\pi  \\
                             \eta\eta_c  
                           \end{array}  &
            \protect\left( \begin{array}{cc} 
                           0 & -2\sqrt{p_\pi p_{\eta_c}} \\
                           -2\sqrt{p_\pi p_{\eta_c}} & 0 \\
            \end{array}\protect\right) \end{array} \times e/\sqrt{2}(4\pi m_{B^0})
\ee
The momenta in the two systems are respectively, $p_\pi \cong (m_{B^0}/2), p_{\eta_c} \cong
0.7 (m_{B^0}/2)$. The single dimensionless parameter $e$ controls the strength of the
effective ``scattering length'', $(\sqrt{2} e)/(4\pi m_{B^0})$, \footnote{For simplicity,
we do not deal with diagonal elements. For elastic scattering, it is the diagonal elements
which are non-zero, and these $K$ matrix elements are simply the negative of the tangent
of the elastic phase shifts. Thus, here the quantity $(\sqrt{2}e/(4\pi m_{B^0})$ is the
magnitude of an effective ``scattering length'' which acts as a measure of the strength
of the two-body quasi-elastic interaction $\pi^+\pi^-\to \eta\eta_c$, at c.~m.~energy
of $m_{B^0}=5.28\GeV$.} which is
taken as $>0$. A priori, a parameter like $e$ is of order unity.\cite{ref12}
We shall allow $e$ to be ``enhanced'', in order to see what is 
necessary to represent the BELLE data. We do not consider an effective ``scattering length''
larger than a few times $(1/m_{B^0})$, that is $<(1/\GeV)$, corresponding to a maximal
strong-interaction phase of about 60$^\circ$ in magnitude. Unitarity is respected by using $1/(1+iK)$ in
the decay amplitudes\cite{ref12}, leading to the Watson strong-interaction phase factors
$e^{i\Delta_{1,2}}$ for the eigenstates. The eigenstates of $K$ are simply 
$|1> = (|\pi^+\pi^-> + |\eta\eta_c>)/\sqrt{2}$ with eigenvalue phase $\Delta_1$ given by
$\tan\Delta_1 = +(\sqrt{2p_\pi p_{\eta_c}}e)/(4\pi m_{B^0})$, and $|2> =
(|\pi^+\pi^-> - |\eta\eta_c>)/\sqrt{2}$ with eigenvalue phase $\Delta_2$ given by
$\tan \Delta_2 = -(\sqrt{2p_\pi p_{\eta_c}}e)/(4\pi m_{B^0})$. Note that there are
the two essential strong-interaction phases, and that these are opposite in sign.
This feature leads to an immediate doubling of the magnitude of the phase
difference $(\Delta_2-\Delta_1)$, and the direct CP-violating asymmetry parameter $C_{\pi\pi}$ is proportional
to $\sin(\Delta_2-\Delta_1)$. 
This doubling
is an important effect obtained in this model, also because it can take individual phases
in the fourth and first quadrants into a phase difference in the third quadrant, with
the resulting negative sign for the cosine and sine of the phase difference. It is the
negative cosine which gives rise to an enhancement of the magnitude of $C_{\pi\pi}$,
in Eq.~(10) below. The state 
$|\pi^+\pi^-> = (|1>+|2>)/\sqrt{2}$; the explicit dependence on the strong-interaction
phases in the decay amplitudes can be exhibited as
\ba
\ol{B}_{\pi\pi} &=& e^{i\Delta_1}\left(A_1+A_2e^{i(\Delta_2-\Delta_1)}\right)\nn\\
B_{\pi\pi} &=& e^{i\Delta_1}\left(A_1^* + A_2^* e^{i(\Delta_2-\Delta_1)}\right)
\ea
The two complex amplitudes $A_{1,2}$ involve the weak interactions; explicitly
the KM phase $\delta$ from the quark mixing matrix. Within this model with eigenstates
$(|\pi^+\pi^-> \pm |\eta\eta_c>)/\sqrt{2}$, we are led to an ansatz for the explicit forms $A_{1,2}$,
\ba
A_1 &=& A e^{-i\delta}(1+a e^{i\delta}) \nn\\
A_2 &=& A e^{-i\delta}(1-a e^{i\delta}) \nn\\
&&A\;{\mathrm{real}}
\ea
with $a\cong (s_2/s_3)f_{c\ol{c}} = \tilde{s} f_{c\ol{c}} \sim \tilde{s}$.
Here $f_{c\ol{c}}$ represents a ``fraction'' for a $c\ol{c}$ component
in the system $\eta\eta_c$; we use $f_{c\ol{c}}\sim 1$ in numerical estimates.
The ratio of the sines of small mixing angles $(s_2/s_3) = \tilde{s} >1 $
(we use $\tilde{s}\sim 2.5$ in numerical estimates\cite{ref14}), appears
as the result of the relative values of different elements of the CKM matrix
(in its original form\cite{ref15}, or as $(s_{12}s_{23}/s_{13})$ in the
``standard'' form\cite{ref14}). The relevant elements involve that for
$b\to c$ times that leading to $\ol{c}d$, relative to that for $b\to u$ times
that leading to $\ol{u}d$. (We approximate the cosines of the small CKM angles
as $\sim 1$.) 
Thus, the amplitudes $A_{1,2}$ involve two different
CP-violating phases $\delta_{1,2}$, as functions of $\delta$.
\ba
A_1 &=& A N_1 e^{-i\delta}e^{i\delta_1};\;\; \tan\delta_1 = 
\frac{\tan\delta}{(1+(1/\tilde{s}\cos{\delta}))}\nn\\
A_2 &=& A N_2 e^{-i\delta}e^{i\delta_2};\;\; \tan\delta_2 = 
\frac{\tan\delta}{(1-(1/\tilde{s}\cos{\delta}))}\nn\\
{\mathrm{with}}\;N_{1,2}&=& \left\{(1\pm \tilde{s}\cos\delta)^2 + (\tilde{s}\sin\delta)^2\right\}^{1/2}
\ea
The phase difference $(\delta_2-\delta_1)$ is not zero unless $\delta\to 0$ (i.~e.~$\delta_1=\delta_2=0$).
Using Eqs.~(6,7) we have, in the model
\ba
\ol{B}_{\pi\pi} &=& A' e^{-i\delta} e^{i(\Delta_1+\delta_1)}\left\{ r+e^{i(\Delta_2-\Delta_1)}e^{i(\delta_2-\delta_1)}\right\}\nn\\
B_{\pi\pi} &=& A' e^{+i\delta} e^{i(\Delta_1-\delta_1)}\left\{ r+e^{i(\Delta_2-\Delta_1)}e^{-i(\delta_2-\delta_1)}\right\}\nn\\
{\mathrm{with}}\; r&=&(N_1/N_2),\;\; A'=AN_2
\ea
Then
\ba
\lambda &=& \left(\frac{q}{p}\right)\left(\frac{\ol{B}_{\pi\pi}}{B_{\pi\pi}}\right)\nn\\
&=& e^{-i(2\beta+2\delta)}e^{i2\delta_1}\left\{
\frac{r+e^{i(\Delta_2-\Delta_1)}e^{i(\delta_2-\delta_1)}}{r+e^{i(\Delta_2-\Delta_1)}e^{-i(\delta_2-\delta_1)}}\right\}
\ea
using\cite{ref13,ref14} $(q/p)=e^{-i2\beta}$, $2\beta\sim 45^\circ$. 
We thus obtain explicit formulae for the 
$C_{\pi\pi}$ and $S_{\pi\pi}$ in Eq.~(2a), using the model; we use\cite{ref14} 
$\delta\sim 45^{o}$ in numerical estimates.
\ba
C_{\pi\pi}&=& +\frac{(\sin \Delta_{21})(\sin\delta_{21})}{\left(\frac{1+r^2}{2r}\right)+(\cos\Delta_{21})(\cos\delta_{21})}\nn\\
S_{\pi\pi}&=& +\cos(2\beta+2\delta-2\delta_1) 
\left\{ \frac{(r\cos \Delta_{21}+\cos\delta_{21})\sin\delta_{21}}
{\left(\frac{1+r^2}{2}\right)+r(\cos\Delta_{21})(\cos\delta_{21})}\right\}\nn\\
&&-\sin(2\beta+2\delta-2\delta_1)
\ea
with $r^2\cong 2.95$; $\tan\delta_1 \cong 1/(1+1/(2.5\times .7)) \cong 0.635 
\Rightarrow \delta_1 \cong 32.5^\circ$; \par\noindent
$\tan\delta_2 \cong 1/(1-1/(2.5\times .7))\cong 2.32 \Rightarrow \delta_2 \cong 66.5^\circ$ 
and thus
$\delta_{21}=(\delta_2-\delta_1)\cong 34^\circ$;
$\Delta_{21}=(\Delta_2-\Delta_1)$.

Clearly, as $\delta\to 0$ we have $\delta_{1,2}\to 0$ and $|\lambda|\to 1$; then
$C_{\pi\pi}\to 0$ and $S_{\pi\pi}\to -\sin 2\beta \sim -0.7$, as measured in 
\cite{ref9,ref10}.
As stated after Eq.~(3), the BELLE value for $A_{\pi\pi} = -C_{\pi\pi}$ is large;
in the context of the above formula this requires a very large $\Delta_{21}$ to obtain
an enhancement factor from the denominator. We consider $\Delta_{21}\cong -110^\circ$.
As stated following Eq.~(4), this is nearly the largest phase difference that we can consider
within the framework of this model. Then from Eqs.~(10), we calculate
\ba
C_{\pi\pi} &\sim& -\frac{(\sin 110^\circ )(\sin 34^\circ)}{1.15+(\cos 110^\circ)(\cos 34^\circ)} 
\sim -0.6\nn\\
S_{\pi\pi} &\sim& (\cos 70^\circ) \left\{ \frac{(1.72 \cos 110^\circ + \cos 34^\circ)
(\sin 34^\circ)}{1.97+ 1.72\cos 110^\circ \cos 34^\circ}\right\}- \sin 70^\circ\nn\\
& & \sim -0.9
\ea
Such a large difference between two strong-interaction phases is a surprising new
result.\footnote{It is interesting that the BELLE collaboration has reached a similar
conclusion about the necessity for a very large strong-interaction phase difference 
(also $\sim -110^\circ$, note Fig.~9 in \cite{ref3}), using different 
phenomenological formulae which originate
in quark dynamics, (but are similar in form to our Eq.~(10)). However, there is no
indication of how such a large phase difference might arise from high-energy strong
interactions involving perturbative gluons and elementary quarks. Neither is it apparent
how hypothetically different, strong-interaction phases for different amplitudes at this
level, are carried over to the state of physical hadrons.} From our discussion following 
Eq.~(4),
\ba
\Delta_{21} &=& \left\{ \tan^{-1}(-\omega) - \tan^{-1}\omega\right\} \cong -110^\circ\nn\\
\Rightarrow  \omega &=&\sqrt{p_\pi p_{\eta_c}}\left(\frac{\sqrt{2}e}{4\pi m_{B_0}}\right)
=\sqrt{p_\pi p_{\eta_c}}{\cal{A}} \sim 1.43
\ea
and thus there is an effective ``scattering length'' of
\be
{\cal{A}} = \left(\frac{\sqrt{2}e}{4\pi m_{B^0}}\right) \sim \frac{1.43}{\sqrt{p_\pi p_{\eta_c}}}
\sim \frac{3.4}{m_{B^0}} \cong 0.65/\GeV < 1/\GeV
\ee
This dynamical model thus allows, in principle, for such a strong quasi-elastic interaction.\footnote{
Extension of the idea of using coupled channels$^{F5}$,
to include $\eta',\eta(1295),\eta(1440)$, could increase the effective strength (as then
roughly summarized by the parameter ${\cal{A}}$ in Eq.~(13)). 
}
However, in
the context of two-body interactions of physical hadrons at 5.28 GeV c.~m.~energy, 
the ${\cal{A}}$ in Eq.~(13) appears rather large. For example, taking\cite{ref12} $e\sim 1$
gives only a small ${\cal{A}} \sim 0.11/m_{B^0} \cong 0.02/\GeV$. Then, $\Delta_{21}=
(\Delta_2-\Delta_1) \sim -2\omega \sim -0.09$, which results in a much smaller parameter
for direct CP violation 
\be
C_{\pi\pi} \sim  -0.025
\ee
If there is a large $C_{\pi\pi}$, a measurable prediction for the decay mode
$\ol{B}^0(B^0)\to \eta\eta_c$ is that $(C_{\eta\eta_c})_{\mathrm{b.r.}}=-(C_{\pi\pi})_{\mathrm{b.r.}}$,
where b.r.~denotes multiplication of $C$ by the corresponding branching ratio. Also, from only
the quasi-elastic strong-interactions which we have discussed here, the parameter
for direct CP violation in $\ol{B}^0(B^0)\to \pi^0\pi^0$ is $(C_{\pi\pi}^0)_{\mathrm{b.r.}} = (C_{\pi\pi})_{\mathrm{b.r.}}$.

The main result of this paper is the sizable value calculated for $A_{\pi\pi}=-C_{\pi\pi}$.
In connection with the present experimental results \cite{ref2,ref3} for $S_{\pi\pi}$,
we note that there is a different simple possibility for the $S_{\pi\pi}$ calculated
within the framework of Eqs.~(10) obtained from this dynamical model. This involves the
possibility within the model of $\delta_{1,2}\rightarrow \delta_{2,1}$ with $N_{1,2}
\rightarrow N_{2,1}$, and $\Delta_{1,2}\rightarrow \Delta_{2,1}$. This illustrates the
role of the first term for $S_{\pi\pi}$ in Eq.~(11) which then gives rise to 
a somewhat smaller 
$-S_{\pi\pi}$. $C_{\pi\pi}$ is unchanged. This suggests that it is not natural
to have a sizable $C_{\pi\pi}$ together with a very small $S_{\pi\pi}$. On the other
hand, if final-state interactions are completely neglegible,  $C_{\pi\pi}\rightarrow 0$;
then we have $S_{\pi\pi}\rightarrow \sin(2\beta+2\delta)$, which could be $\sim 0$ if
$(2\beta+2\delta)$ is $\sim 180^\circ$.

In summary, we have shown that the large, direct CP-violating parameter $|C_{\pi\pi}|$
reported\cite{ref1,ref3} by the BELLE collaboration implies the presence of a very
large difference between (at least) two strong-interaction phases. We have shown
how strong, two-body quasi-elastic interactions between physical hadrons at $m_{B^0}=5.28$
GeV, can accommodate such an unusual strong-interaction phase difference.$^{F8}$
This type of model might allow new possibilities for estimating the strong-interaction
phases which are necessary for observable direct CP violation, in other two-body
final states from $\ol{B}^0(B^0)$ decay.

S.~B.~thanks Lahlit Sehgal and Kazuo Abe for information.
 
\section*{Added note}
The BABAR collaboration has just reported the observation of the decays
$B^\mp \to \pi^\mp\eta$, and an interesting negative, direct CP-violating
asymmetry (in hep-ex/0303039). In \cite{ref12}, minimal estimates of the
strong-interaction phases are used in estimating the CP-violating 
asymmetries. Values of the $K$-matrix elements for the final-state
strong interactions are taken, which are minimal in magnitude. These are
likely to be larger, by a factor of at least 2. The calculated asymmetries
are multiplied by this factor. S.~B. thanks Janice Button-Schafer for helpful
communications about the new BABAR results.


\begin{thebibliography}{99} 
\bibitem{ref1} BELLE Collab., K.~Abe et al., Phys.~Rev.~Lett.{\bf 89} (2002) 071801
\bibitem{ref2} BABAR Collab., B.~Aubert et al., Phys.~Rev.~Lett {\bf 89} (2002) 281802
\bibitem{ref3} BELLE Collab., K.~Abe et al., hep-ex/0301032, Feb.~2003
\bibitem{ref4} ARGUS Collab., H.~Albrecht et al., Phys.~Lett.~{\bf B192} (1987) 245
\bibitem{ref5} J.~H.~Christenson, J.~W.~Cronin, V.~L.~Fitch and R.~Turlay, Phys.~Rev.~Lett.~{\bf 13} (1964) 138
\bibitem{ref6} NA 31 Collab., H.~Burkhardt et al., Phys.~Lett.~{\bf B206} (1988) 168
\bibitem{ref7} N48 Collab., V.~Fanti et al., Phys.~Lett.~{\bf B465} (1999) 335
\bibitem{ref8} KTeV Collab., A.~Alavi-Harati et al., Phys.~Rev.~Lett.~{\bf 83} (1999) 22 
\bibitem{ref9} BABAR Collab., B.~Aubert et al., Phys.~Rev.~Lett.~{\bf 89} (2001) 091801\\
BABAR Collab., B.~Aubert et al., hep-ex/0203007
\bibitem{ref10} BELLE Collab., K.~Abe et al., Phys.~Rev.~Lett.~{\bf 87} (2001) 091802\\
BELLE Collab., K.~Abe et al., hep-ex/0202027v2
\bibitem{ref11} S.~Barshay and J.~Geris, Phys.~Lett.~{\bf B84} (1979) 319
\bibitem{ref12} S.~Barshay, D.~Rein and L.~Sehgal, Phys.~Lett.~{\bf B259} (1991) 475 
\bibitem{ref13} Y.~Nir and H.~R.~Quigg, Ann.~Rev.~Nucl.~Part.~Sci. {\bf 42} (1992) 211
\bibitem{ref14} Particle Physics Booklet, July 2002, pages 165-172\\
Particle Data Group, K.~Hagiwara et al., Phys.~Rev.~{\bf D66} (2002) 010001-1
\bibitem{ref15} M.~Kobayashi and J.~Maskawa, Prog.~Theor.~Phys.~{\bf 49} (1973) 652
\end{thebibliography}
\end{document}